\documentstyle[aps,graphics]{revtex}

\begin{document}

\title{Nonequilibrium Translational-Rotational Effects in Nucleation }

\author{D. Reguera, J.M. Rub\'{\i}}

\address{Departament de F\'{\i}sica Fonamental, Facultat de F\'{\i}sica,\\
 Universitat de Barcelona, Diagonal 647, 08028 Barcelona, Spain}

\maketitle
\begin{abstract}
The role that translational-rotational degrees of freedom play in nucleation
theories is reconsidered by the introduction of a new formalism that properly
accounts for the effects of motion of clusters in nucleation rate. The analysis
of the non-equilibrium kinetics of the process, performed by retaining the dynamics
of the clusters, enables one to clarify some of the paradoxical aspects that
the inclusion of these degrees of freedom has presented. 
\end{abstract}

\section{Introduction}

In spite of years of investigations and of the fact of being a ubiquitous phenomenon,
nucleation is still a puzzling process not completely understood \cite{kn:oxtoby,kn:debene}.
During the last decade, the implementation of new experiments measuring actual
nucleation rates in liquid-gas systems has revealed the shortcomings of Classical
Nucleation Theory (CNT)\cite{kn:oxtoby}. This fact has impelled the sprouting
of new theories and modifications to CNT attempting to reproduce the experimental
data. The corrections and improvements proposed in the literature are diverse
and not always accepted with unanimity. Examples of these corrections are based
on\cite{kn:wu}: self-consistency considerations; contribution of translational,
rotational and vibrational degrees of freedom; curvature effects in the surface
tension; monomer depletion and cluster scavenging\cite{kn:seinfeld2} or even
the inclusion of effects associated to spatial \cite{kn:seinfeld3} or temperature
inhomogeneities\cite{kn:seinfeld1}.

A very controversial issue in nucleation theory is the proper accounting for
embryo degrees of freedom, sometimes referred to as the translational-rotational
paradox. The controversy dates back to the consideration of Lothe and Pound
\cite{kn:lothe} of what appears to be a serious inconsistency in the conventional
theory of nucleation from the vapor phase. These authors pointed out that several
important contributions to the free energy of formation of the critical cluster
attributable to rotational and translational degrees of freedom had been neglected
in CNT. The inclusion of these additional terms in the way they proposed increased
the nucleation rate by a factor on the order of \( 10^{17} \), enough to destroy
reasonable agreement between theory and experiments.

Reiss and co-workers \cite{kn:rk,kn:rkc} attempted to resolve the paradox by
arguing that experimental surface tension in the capillarity approximation already
takes into account the rotational and most of translational effects. From this
point of view, Reiss-Katz-Cohen theory presents a new way to incorporate these
effects into CNT, leading to a much smaller change in the expected rates, between
\( 10^{3} \) and \( 10^{6} \).

There followed a series of counter arguments and discussions by several authors
\cite{kn:controversia}. Although there has been a long standing controversy,
the issue was never satisfactorily resolved and the proper inclusion of these
contributions continued being debated \cite{kn:ruth}. The issue seems to have
come to an end with the recent work of Reiss, Kegel and Katz \cite{kn:reissprl}
which apparently constitutes the proper resolution of the paradox and clarifies
other inconsistencies of the nucleation theory. 

As we have seen, considerable emphasis has been put on the introduction of the
translational and rotational degrees of freedom in the equilibrium partition
function of a nucleating droplet. However, these corrections refer only to the
influence of these degrees of freedom in the equilibrium sense. Translational
and rotational degrees of freedom also arise from the {\it  motion} and {\it  rotation}
of the clusters through the metastable phase. This is a purely non-equilibrium
effect, which may have a relevant influence in the nucleation kinetics.

Our purpose in this paper is precisely to present a new theory to consistently
incorporate translational and rotational contributions {\it  due to cluster
motion} in nucleation. Contrarily to the theories proposed up to now, we will
adopt a non-equilibrium description more proper for a problem intrinsically
out of equilibrium as nucleation is. In this sense, we will show that when considering
nucleation retaining the dynamics of the problem a new and important correction
to the nucleation kinetics arises. 

The paper is distributed as follows. In Sec. II we analyze the influence of
the translational/rotational degrees of freedom in the nucleation process and
review how these degrees of freedom have been considered from the equilibrium
point of view. To analyze the potential non-equilibrium effects of the motion
of the clusters in nucleation, we derive in Sec. III, the equation governing
the kinetics of the nucleation, retaining the dynamics of the clusters. After
a proper elimination of the velocity variables, this equation will be used in
Sec. IV to obtain a nonequilibrium correction to the nucleation rate, discussing
explicitly the results for nucleation rates of water. Finally, in the Conclusions,
we analyze and summarize the main consequences of our work.

\section{Influence of translational-rotational degrees of freedom in nucleation}

Phenomenological theories of nucleation are based on the formulation of the
free energy \( G(n) \) associated to the formation of a cluster of \( n \)
molecules from the metastable phase. The expression proposed by CNT, based on
capillarity approximation, is obtained by treating the cluster as a macroscopic
spherical droplet with bulk and surface free energy contributions

\begin{equation}
\label{eq:CNT}
G(n)=-k_{B}Tn\ln S+\sigma s_{1}n^{2/3}
\end{equation}
 Here \( T \) is the temperature, \( S \) the saturation ratio, \( \sigma  \)
the surface tension, \( s_{1} \) the mean surface of a single molecule of the
liquid phase and \( k_{B} \) is the Boltzmann constant. However, treating embryos
as macroscopic objects and using thermodynamic arguments to calculate the work
of embryo formation give rise to important inconsistencies \cite{kn:oxtoby,kn:wu}.
Capillarity approximation leave out translational, rotational and configurational
contributions; it predicts a non-zero value for the energy of monomer formation
(\( G(1)\neq 0 \)) and ignores possible dependencies of surface tension on
the temperature or the radius of the clusters. In spite of all of these inconsistencies,
the results predicted by CNT are in most cases in reasonable agreement with
experimental data. Paradoxically, it seems that most of the corrections originally
introduced to restore the logical consistency of CNT spoil this agreement and
leads to worst results. 

Basically, the corrections proposed by these theories in liquid-vapor systems,
can be recovered from the general formula \cite{kn:kiang}

\begin{equation}
\label{eq:fisher}
\frac{G(n)}{k_{B}T}=-n\ln S+k_{n}\frac{\sigma s_{1}}{k_{B}T}n^{\gamma }+\tau \ln n+\ln (Vq_{0})
\end{equation}
 where \( V \) is the volume, and \( k_{n,}\gamma  \), \( \tau  \) and \( q_{0} \)
are parameters whose expressions and values are different for each theory. In
this expression, the first and second terms are the bulk and surface contributions,
respectively, the two contributions considered in CNT. The remaining terms include
rotational, translational, vibrational, configurational, and replacement contributions
related to additional degrees of freedom which capillarity approximation may
leave out. Our analysis will be focused on translational-rotational aspects
of nucleation.

Capillarity approximation deals with the free energy required to form a single
cluster \( at \) \( rest \) in the metastable phase. However, the conception
of droplets at rest constitutes an approximation to the real behavior of the
system. The small clusters which are the embryos of the new phase can spontaneously
appear at any point of the system, and with arbitrary orientation. Moreover,
due to the mesoscopic size of these entities, they {\it  move} and {\it  spin
around} the metastable phase because the influence of the medium in which they
are embedded (Brownian motion). Both factors constitute translational and rotational
degrees of freedom of the cluster and must be taken into account to describe
accurately the nucleation process. But the influence and the way of considering
both effects is different. The arbitrariness of positions and orientations which
a cluster may occupy can be incorporated as a purely equilibrium correction
by including these translational and rotational degrees of freedom in the partition
function of the cluster. Consequently, this yields a modification of the free
energy of formation of this cluster. Contrarily, the effect of the movement
and rotation of the clusters does not alter the nucleation barrier itself. It
is a nonequilibrium factor that modifies the kinetics of the process.

Most of the work on translational-rotational correction, and the paradox itself,
addresses to modify the free energy to form a cluster of size \( n \), taking
into account these degrees of freedom in the partition function. In this sense,
one may refer to the works of Lothe and Pound (LP)\cite{kn:lothe}, and the
more recent and accurate theory of Reiss, Kegel and Katz (RKK)\cite{kn:reissprl}.
These works deal with purely equilibrium corrections. The common underlying
idea is that if nucleation occurs in a volume V, the physical critical cluster
may appear anywhere in the system with any orientation. Considering that all
these positions of the cluster are equivalent, one then has this additional
translational and rotational freedom to account for in the partition function.
The accounting of this degrees of freedom increases the value of the partition
function, consequently reducing the free energy of formation of a droplet and
thus increasing the nucleation rate. 

The difference between both treatments and the origin of the paradox concerns
what is the proper accounting of this degrees of freedom in the partition function
and how to define and distinguish different states of a cluster in a volume
V. Using the CNT expression for the free energy of formation of a cluster entails
the extrapolation of a macroscopic expression to the mesoscopic scale. Whereas
it is clear how to locate one macroscopic droplet at rest, the characterization
becomes less obvious when this cluster only contains a small number of molecules.
For instance, the position of a mesoscopic cluster constituted by \( n \) molecules
inside a spherical volume \( v \), can be described in terms of the position
of its center of mass, or by the position of the center of the spherical container.
Whereas for a macroscopic droplet both description practically coincide, for
the mesoscopic one they differ due to the fluctuations. 

The other problem arising in the mesoscopic scale concerns the pertinent volume
scale one has to use to discretize the space with the purpose of distinguishing
different states in the calculation of the partition function. The uncertainty
in defining a volume scale is a well-know problem in Statistical Mechanics \cite{kn:hill}.
For macroscopic systems, however, this arbitrariness of the length scale is
not relevant, because only introduces a logarithmic correction to the free energy
which is always negligible in the thermodynamic limit. But for a small embryo,
the correction may become important. Hence the problem and the central point
in the paradox is to know what is the proper and physical meaningful volume
scale to count states in the partition function and consequently to calculate
the free energy of a mesoscopic cluster.

A reasonable choice for the volume scale would be of course \( \Lambda ^{3} \),
based on the de Broglie wavelength scale \( \Lambda  \); that was the one used
by Lothe and Pound \cite{kn:lothe}. They assumed that the location of a spherical
cluster of \( n \) molecules is defined through the position of its center
of mass and its orientation. Therefore every configuration of \( n \) molecules
inside a spherical volume \( v \) with different position of the center of
mass and different orientation constitute a different realization of the same
cluster and consequently a new state to be accounted for in the partition function.
The minimum uncertainty to locate and distinguish different states is thus the
quantum uncertainty. That is, every configuration of \( n \) molecules whose
center of mass and orientation differs by \( \Lambda  \) and \( \Lambda _{rot} \),
respectively is a new state that increases the value of the partition function.
This criterion leads to the following correction of the free energy

\noindent \begin{eqnarray}
\Delta G_{LP}(n)=-k_{B}T\ln \left( \frac{V}{\Lambda ^{3}}\frac{8\pi ^{2}}{\Lambda _{rot}^{3}}\frac{1}{q_{rep}}\right)  & \label{eq:LP} 
\end{eqnarray}
 where \( \Lambda =\left( \frac{h^{2}}{2\pi m_{1}k_{B}Tn}\right) ^{1/2} \),
and \( \Lambda _{rot}=\left( \frac{h^{2}}{2\pi Ik_{B}T}\right) ^{1/2} \), with
\( m_{1} \) the mass of a single monomer, \( I \) the moment of inertia of
the cluster, \( h \) the Planck's constant, and \( q_{rep} \) a {}``replacement
factor{}'' related to the entropy reduction accompanying the separation of
\( n \) molecules from the system. Assuming reasonable values for vapor condensation,
the small size of the de Broglie lengths, originates that these additional terms
increase nucleation rates by approximately \( 10^{17} \), destroying in most
cases the agreement between CNT results and experimental data.

But the de Broglie length is not the only reasonable scale one may choose, and
RKK theory \cite{kn:reissprl} proposes a different alternative. From their
point of view, the location of a cluster must be defined by the position of
the spherical container, and not by its center of mass. All the possible configurations
of \( n \) molecules inside a spherical container of volume \( v \) correspond
then to the {\it  same} cluster and consequently are already accounted for in
the CNT expression of the free energy. In particular, all the possible orientations
of the molecules inside the spherical volume are included, which implies that
the rotational degrees of freedom are already accounted for in the CNT free
energy barrier. In addition, to incorporate the translational degrees of freedom
one has to take into account that not all the different positions which this
spherical cluster may occupy inside the volume V correspond to different clusters.
If we displace the spherical container a small distance \( dx \), most of the
configurations of the \( n \) molecules compatible with the new situation are
the same and have been already accounted for in the undisplaced original cluster.
Hence treating this new position of the cluster as a different cluster entails
an overcounting of states in the partition function. They affirm that this is
essentially the origin of the excessively high correction of obtained by LP,
the overcounting of configurations in choosing the quantum length scale to count
states in the partition function. Consequently, there exist a minimum volume
scale to differentiate clusters in the configurational space. For a nearly incompressible
drop, this scale turns out to be the one related to the volume fluctuation \( \sqrt{k_{B}T\kappa v_{1}n} \).
That is, a cluster cannot be located more precisely than its volume fluctuation.
This criteria prevents the overcounting of configuration and gives rise to a
correction in the nucleation barrier\[
\Delta G_{RKK}=-k_{B}T\ln \frac{V}{\sqrt{k_{B}T\kappa v_{1}n}}\]
leading to a much smaller change in nucleation rates of the order of \( 10^{4} \).

Both LP and RKK approaches are deeply rooted in equilibrium statistical arguments.
However, the correction they propose does not in fact concern the {\it  movement}
of clusters. Different ways exist by which this movement can influence the process. 

On one hand, the movement of the clusters through the metastable phase may alter
the rate at which the cluster incorporates molecules of the metastable phase.
Olson and Hamill \cite{kn:olson} considered the effect of the movement of clusters
in the supersaturated vapor in the rate at which a cluster gains monomers \( \beta (n) \).
Yet, the correction to \( \beta (n) \) has been shown to be negligible\cite{kn:wu}. 

But on the other hand, the movement of the cluster can modify not only the rate
of growth of a cluster, but also the kinetics of the whole process. Contrarily
to the equilibrium point of view, our objective will be to analyze the influence
of the {\it  movement} and {\it  rotation} of clusters in the nucleation process
from a dynamical point of view. Nucleation is a nonequilibrium kinetic phenomenon
essentially related to the variation of the size distribution of clusters present
in the system. If the clusters are moving through the metastable phase, its
movement may influence the evolution of the cluster size distribution thus altering
the kinetics of the process. Therefore, it seems reasonable to study the process
not only by considering the evolution of the size of clusters but also retaining
its dynamics. This is the approach we develop in the next section.

\section{Nonequilibrium kinetics of nucleation}

We will model the real system, composed by droplets and molecules of the metastable
phase, as a dispersion of clusters of different and varying sizes. Since we
will focus on homogeneous nucleation in spatial homogeneous systems, the spatial
variables are not relevant. Thus, the variables characterizing a cluster are
its size \( n \), its velocity \( \vec{u} \) and its angular velocity \( \vec{\omega } \)
. The description of the system will be carried out in terms of the distribution
function \( f(\underline{\Gamma },t) \), with \( \underline{\Gamma }\equiv (n,\vec{u},\vec{\omega }) \),
whose evolution is dictated by the continuity equation

\begin{equation}
\label{eq:conti}
\frac{\partial f(\underline{\Gamma },t)}{\partial t}=-\frac{\partial }{\partial \underline{\Gamma }}\cdot \underline{J}(\underline{\Gamma },t)
\end{equation}
 where \( \underline{J}\equiv (J_{n},\vec{J}_{u},\vec{J}_{\omega }) \) is the
current of cluster sizes and velocities defined in \( \underline{\Gamma } \)-space.

The nucleation process may be conceived as a diffusion process in the space
spanned by the values of \( \underline{\Gamma } \) through the energy barrier

\begin{equation}
\label{eq:barrera}
C(\underline{\Gamma })=\Phi (n)+\frac{1}{2}m(n)u^{2}+\frac{1}{2}I(n)\omega ^{2},
\end{equation}

\noindent where \( \Phi (n) \) represents the energy of formation of a \( n \)-cluster
\( at \) \( rest \) of mass \( m(n)=nm_{1} \) and moment of inertia \( I(n) \).
For a spherical rigid cluster \( I(n)=\frac{2}{5}m(n)r^{2} \), thus implying
\( I(n)\sim n^{5/3} \). The remaining terms correspond to translational and
rotational kinetic energies.

The entropy of the system, \( S(t) \), is given through the Gibbs entropy postulate
\cite{kn:mazur,kn:klp}

\begin{equation}
\label{eq:Gpostulate}
S(t)=-k_{B}\int f(\underline{\Gamma },t)\ln \frac{f(\underline{\Gamma },t)}{f_{eq}(\underline{\Gamma })}d\underline{\Gamma }+S_{eq}
\end{equation}
where \( S_{eq} \) is the value of the entropy at equilibrium. Its variations
can be expressed in the form

\begin{equation}
\label{eq:dS}
\delta S(t)=-\frac{1}{T}\int \mu (\underline{\Gamma },t)\delta f(\underline{\Gamma },t)d\underline{\Gamma }
\end{equation}
 where \( \mu (\underline{\Gamma },t) \) is a generalized chemical potential
defined in \( \underline{\Gamma } \)-space.

\begin{equation}
\label{eq:mu}
\mu (\underline{\Gamma },t)=k_{B}T\ln \frac{f(\underline{\Gamma },t)}{f_{eq}(\underline{\Gamma })}+\mu _{eq}
\end{equation}

\noindent This expression corresponds to the chemical potential of an ideal,
i.e. non-interacting, dispersion of clusters in the liquid phase. In the latter
equation, \( f_{eq}(\underline{\Gamma }) \) represents the equilibrium distribution
given by

\begin{equation}
\label{eq:feq}
f_{eq}(\underline{\Gamma })\propto \exp (-\frac{C(\underline{\Gamma })}{k_{B}T})
\end{equation}

\noindent and \( \mu _{eq} \) is the chemical potential at equilibrium.

The expression for the current \( \underline{J} \) defined in Eq. \ref{eq:conti}
can be obtained from nonequilibrium thermodynamics. The entropy production \begin{equation}
\label{eq:sigma}
\sigma =-k_{B}\int \underline{J}\cdot \frac{\partial }{\partial \underline{\Gamma }}\ln \frac{f}{f_{eq}}d\underline{\Gamma }
\end{equation}

\noindent follows from Eqs. (\ref{eq:conti}), (\ref{eq:dS}) and (\ref{eq:mu}).
The resulting linear law, obtained from the assumption of isotropy and locality
in \( \underline{\Gamma } \)-space \cite{kn:degroot}, is given by 

\noindent \begin{equation}
\label{eq:linear_law}
\underline{J}(\underline{\Gamma },t)=-\underline{L}\cdot \frac{\partial }{\partial \underline{\Gamma }}\ln \frac{f}{f_{eq}}
\end{equation}
where \( \underline{L} \) is the corresponding matrix of Onsager coefficients.
This expression can be used in the continuity equation (\ref{eq:conti}) thus
leading to the Fokker-Planck equation

\begin{equation}
\label{eq:fp}
\frac{\partial f}{\partial t}=\frac{\partial }{\partial \underline{\Gamma }}\cdot \left( \underline{D}\cdot \frac{\partial f}{\partial \underline{\Gamma }}+\frac{\underline{D}}{k_{B}T}\cdot \frac{\partial C}{\partial \underline{\Gamma }}f\right) 
\end{equation}
 where \( \underline{D}=\frac{\underline{L}}{Tf} \) is the diagonal matrix
of diffusion coefficients.

The former equation provides a complete dynamical description of the system
in terms of the variables \( n \), \( \vec{u} \) and \( \vec{\omega } \).
However, from the analysis of the time scales of the system it is easy to realize
that in the nucleation time scale the system reaches equilibration in velocities
space. The characteristic nucleation relaxation time\cite{kn:seinfeld1} is
given by \( \tau _{N}\approx \frac{1}{\beta (n)} \) where for the rate \( \beta (n) \)
of addition one molecule to a droplet of size \( i \) we may adopt the classical
expression taken from kinetic theory of gases \( \beta (n)=\frac{p}{\sqrt{2\pi m_{1}kT}}s_{1}n^{2/3}. \)

Similarly, the Brownian translational and rotational time scales are \( \tau _{trans}\approx \frac{m}{6\pi \eta a} \)
and \( \tau _{rot}\approx \frac{I}{8\pi \eta a^{3}} \) respectively , where
\( m \) is the mass of the Brownian particle, \( a \) is its radius and \( \eta  \)
is the viscosity. For nucleation in liquid-vapor systems, the order of magnitude
of these characteristic times is approximately \( \tau _{N}\approx 10^{-8}n^{2/3} \)
and \( \tau _{trans},\: \tau _{rot}\approx 10^{-13}n^{2/3} \). Therefore a
clear separation between time scales exists, the condition \( \tau _{N}\gg \tau _{trans},\: \tau _{rot} \)
holds, and one can perform an adiabatic elimination of the fast variables (the
velocities) \cite{kn:Risken}, assuming that they reach their equilibrium distribution.

After performing the adiabatic elimination of velocities, the resulting dynamics
is governed by the Fokker-Planck equation

\begin{eqnarray}
\frac{\partial f(n,t)}{\partial t}=\frac{\partial }{\partial n} &  & \left( \bar{D}_{N}(n)\frac{\partial f}{\partial n}+\right. \label{eq:fpa} \\
 &  & \; \; \; \; \left. +\frac{\bar{D}_{N}(n)}{k_{B}T}\left( \frac{d\Phi (n)}{dn}+4k_{B}T\frac{1}{n}\right) f\right) \nonumber 
\end{eqnarray}
 We may now identify the diffusion coefficient \( \bar{D}_{N}(n) \) with the
forward rate \( \beta (n) \), and obtain the expression for the effective nucleation
barrier \( \Delta G(n) \)

\begin{equation}
\label{eq:barri}
\Delta G(n)=\Phi (n)+4k_{B}T\ln n
\end{equation}
 The conclusion is that translational and rotational motion of clusters in the
spatial homogeneous medium introduces an {\it  effective} additional size-dependent
contribution \( 4k_{B}T\ln n \) in the nucleation barrier, inherent to the
diffusion process and thus independent of the energy of formation of cluster
at rest \( \Phi (n) \). It is important to highlight that since this contribution
is always positive translational-rotational motion always increases nucleation
barrier and consequently reduces nucleation rates, contrarily to the results
of the equilibrium corrections which always decrease the nucleation barrier
thus increasing the rate. One may wonder about the reasons for this feature.
In essence, the difference lies on the treatment of velocity variables.

From the equilibrium point of view, it was always assumed that the velocity
and the motion of the clusters does not influence the nucleation process. And
the justification was precisely the same that the one we have used to eliminate
the velocity variables in our Fokker-Planck equation. The time scale of the
velocity relaxation is much faster than the nucleation time scale, so it was
assumed that nucleation occurs when velocity distribution is equilibrated thus
not influencing the process. Following this line of reasoning, the free energy
of formation of clusters is constructed by averaging the contribution of all
cluster velocities. The nucleation kinetics is then studied focusing only on
the evolution of the cluster size distribution.

In contrast with these lines of reasoning, in our treatment we consider that,
due to the coupling between the dynamics of \( n \) and (\( \vec{u} \), \( \vec{\omega } \)),
the elimination of velocities must be performed at the last stage because it
influences the kinetics of the whole process. Thus we maintain velocities of
a cluster as relevant variables in the description of the process. Instead of
constructing an averaged free energy of a drop taking into account all possible
velocities, we directly use the free energy to form a \( n \)-cluster with
velocities \( \vec{u} \), \( \vec{\omega } \) (Eq. \ref{eq:barrera}) to study
the evolution of the system as well with respect to the size of clusters as
their velocities. Finally, we perform an adiabatic elimination of \( \vec{u} \)
and \( \vec{\omega } \) from characteristic time scales considerations. The
result is a net \( positive \) contribution to the nucleation barrier. 

In both cases, the velocity is finally eliminated using the same equilibrium
distribution and the same arguments. One then may wonder what is the physical
origin of the additional effective contribution we obtain. The underlying reason
can be naively understood as follows. Although the velocity is initially equilibrated
and equilibrates very fast, every time that a cluster gains or looses a molecule
not only the size distribution of the clusters is altered, but also the velocity
distribution.Therefore, the alteration of cluster size distribution entails
a reorganization of velocities distribution through diffusing currents. These
currents involve an additional dissipation which has an energetic cost. Consequently,
the growing of the cluster requires more effective energy, because part of that
is expended in the reorganization (in the new equilibration) of the velocity
distribution. As we will show in the following section, the nucleation rate
is thus slowed down.

\section{Nucleation rate}

The consideration of translational and rotational contribution due to the motion
of the clusters in the nucleation process then leads to the following expression
for the stationary nucleation rate \cite{kn:McD}, derived from Eq. (\ref{eq:fpa})
\begin{equation}
\label{eq:rate1}
J=A^{*}\exp \left( -\frac{\Delta G(n^{*})}{k_{B}T}\right) 
\end{equation}
 where \( n^{*} \) is the size of the critical nucleus obtained from the condition
of maximum of the nucleation barrier

\begin{equation}
\label{eq:icrit}
\Phi ^{\prime }(n^{*})+4k_{B}T\frac{1}{n^{*}}=0,
\end{equation}
 and \( A^{*} \) is the pre-exponential factor which for liquid-gas nucleation
is given by

\begin{equation}
\label{eq:preexp}
A^{*}=\beta (n^{*})\frac{1}{V}\sqrt{-\frac{1}{2\pi k_{B}T}\frac{\partial ^{2}\Delta G(n^{*})}{\partial n^{2}}}
\end{equation}

In order to quantitatively illustrate the effect of our correction in the nucleation
rate, we will particularize our general formulation to a concrete physical situation
by using a specific model for the energy barrier \( \Phi (n) \) of formation
of clusters at rest. Our aim in this sense is to show how the consideration
of the translational-rotational degrees of freedom in the way we propose may
lead some phenomenological theories to a better agreement with experimental
data for some substances.

As a concrete example we will focus on the nucleation rates of water, comparing
with the experimental results of Viisanen et. al. \cite{kn:reiss}. As a model
for the free energy of formation of a cluster at rest \( \Phi (n) \), we will
use RKK model, explicitly 

\[
\Phi _{RKK}(n)=-k_{B}Tn\ln S+\sigma s_{1}n^{2/3}-k_{B}T\ln \frac{V}{\sqrt{k_{B}T\kappa v_{1}n}}\]
Data for the compressibility of supercooled water are taken from Ref.\cite{kn:angel}.
As we have discussed in Section II, the RKK model constitutes an improvement
of the CNT in the sense that solves some consistency problems and properly incorporates
the equilibrium corrections due to the translational degrees of freedom. Unfortunately,
this theoretical consistency does not guarantee an improvement of the quantitative
results. In fact, for the particular case of water, the RKK theory clearly overestimates
the nucleation rate by factor of \( 10^{5}-10^{6} \), whereas the discrepancy
of the CNT is approximately two orders of magnitude (see Fig. 1).  

However, taking into account the nonequilibrium effects of the motion of the
cluster in the way we propose and using the RKK energy barrier one can achieve
a very good concordance with the experimental results. In Fig. 1, we compare
the experimental values of Ref. \cite{kn:reiss} for the nucleation rate of
water at different temperatures, with the predictions of CNT, and the values
obtained with our theory from Eq. \ref{eq:rate1}. We can see that the agreement
of our result is excellent, specially at high temperatures. For low temperatures,
the results get slightly worst probably because the low temperature anomalies
of the compressibility of the supercooled water (in fact it diverges at approximately
228 K, see Ref \cite{kn:angel}). Nevertheless, the maximum discrepancies are
less than a factor of 3 for the whole range of temperatures above 228 K. This
example illustrates the fact that an energy barrier constructed under the requirement
of preserving self-consistency and including properly the nonequilibrium effects
may reproduce experimental results.

It is important to emphasize that the accuracy of the results we may provide
mostly relies on the correctness of the energy barrier at rest \( \Phi (n) \)
we adopt. The theory we present aims to analyze the nonequilibrium effects of
the motion of the clusters in nucleation, but phenomenological nucleation barriers
based on capillarity approximation present other serious drawbacks , like self-consistency
considerations, or dependencies of the surface tension on the radius of cluster
or temperature, on which agreement with experiments depends on. It is evident,
that in this cases translational corrections are not sufficiently important
to reestablish in those cases agreement with experiments.

On what concerns nucleation in liquids, there is a general belief that cluster
motion is of little influence and does not need to be considered. Homogeneous
nucleation of crystals is much less well understood than condensation discussed
previously, and presents additional difficulties\cite{kn:kelton}. The first
one is related with the lack of a kinetic approach to estimate the pre-exponential
factor \( A^{*} \) or even the forward rate \( \beta (n) \). The second difficulty
concerns the absence of independent measurements of the solid-liquid surface
free energy in the undercooled regime. In fact, the surface free energy is mostly
evaluated from nucleation rate experiments as a fitting parameter. Thus, we
cannot perform direct tests of nucleation rate experiments against crystal nucleation
theories.

However, in spite of the differences between gas and crystal nucleation, and
the additional problems that the latter presents, the formalism previously developed
to introduce translation-rotation effects in liquid-gas nucleation remains applicable.
Consequently, the net effect of translational-rotational degrees of freedom
predicted by our model is again the variation of the energy barrier given in
Eq. (\ref{eq:barri}). In this case, the differences between our nonequilibrium
treatment and the equilibrium approaches become even more obvious, because in
condensed phases translational/rotational contributions to the free energy of
a cluster (in the partition function) are expected to be negligible\cite{kn:kelton},
while the influence of the movement of clusters in the kinetics of the process
still remains the same.

\section{Conclusions}

In summary, in this paper we have reconsidered the role played by the translational-rotational
degrees of freedom through the introduction of a formalism that properly accounts
for the effects of motion of clusters in nucleation rate. The analysis of the
problem has been performed by taking into account the fact that the emerging
clusters remain embedded in the metastable phase and that their movement could
induce modifications in the nucleation rate. Actually, the motion of the clusters
influence the kinetics of size evolution of clusters, therefore the study of
the problem must be performed retaining their dynamics. The influence of the
medium is then reflected in the appearance of a new contribution to the energy
barrier which is independent of the definition of cluster and of the energy
barrier at rest one adopts. The contribution we have found originates from Brownian
diffusion of clusters and in this sense is a translational contribution arising
from a nonequilibrium situation and not from equilibrium statistical considerations.
The effect we have introduced is fundamental as is due to the ineluctable presence
of the medium during the nucleation process. 

The conclusion is that the motion of the clusters is relevant in the kinetics
of nucleation, leads to a reduction of the nucleation rates, and its proper
inclusion constitutes a first step toward the construction of a nucleation energy
barrier that preserve all logical consistencies and is able to reproduce and
predict experimental results.

The revision we propose is unable by itself to correct all possible disagreements
between theories and experimental results. The field is still open, and more
investigations are needed, specially the ones aimed to find out a better expression
for the energy of formation of clusters at rest. When this objective is achieved,
it still will be essential to include the influence of the medium in order to
obtain a prediction suitable to be contrasted with experiments. The presence
of the medium should then be taken into account in future developments of nucleation
theories.

A methodological aspect to be emphasized is that our analysis, leading to kinetic
equations of the Fokker-Planck type, could be directly extended to consider
other hydrodynamic effects in nucleation processes, as the ones related to the
presence of gradients or inhomogeneities. We then provide a theoretical framework
from which the influence that the proper dynamics of the metastable phase may
play in the nucleation process could be studied systematically.

\acknowledgements We would like to thank T. Alarc\'{o}n, A. P\'{e}rez-Madrid,
H. Reiss, R. Bowles and Y. Djikaev for fruitful discussions, and R. Strey for
sending us their revised experimental data. This work has been supported by
DGICYT of the Spanish Government under grant PB98-1258. D. Reguera wishes to
thank Generalitat de Catalunya for financial support.

\newpage

\vspace{3cm}

\par\centering {}{\large FIGURE CAPTIONS \par{}}{\large \par}

\begin{itemize}
\item {\large Figure 1.- Comparison of experimental nucleation rates \( J \) as a
function of the supersaturation \( S \) with the predictions of the Classical
Nucleation Theory (dashed lines) and the new theory (full lines). Crosses represent
data of Viisanen \( et. \) \( al. \) \cite{kn:reiss} for water at different
nucleation temperatures. \par{}}{\large \par}
\end{itemize}
\newpage

\begin{figure}
{\par\centering \resizebox*{1\columnwidth}{!}{\includegraphics{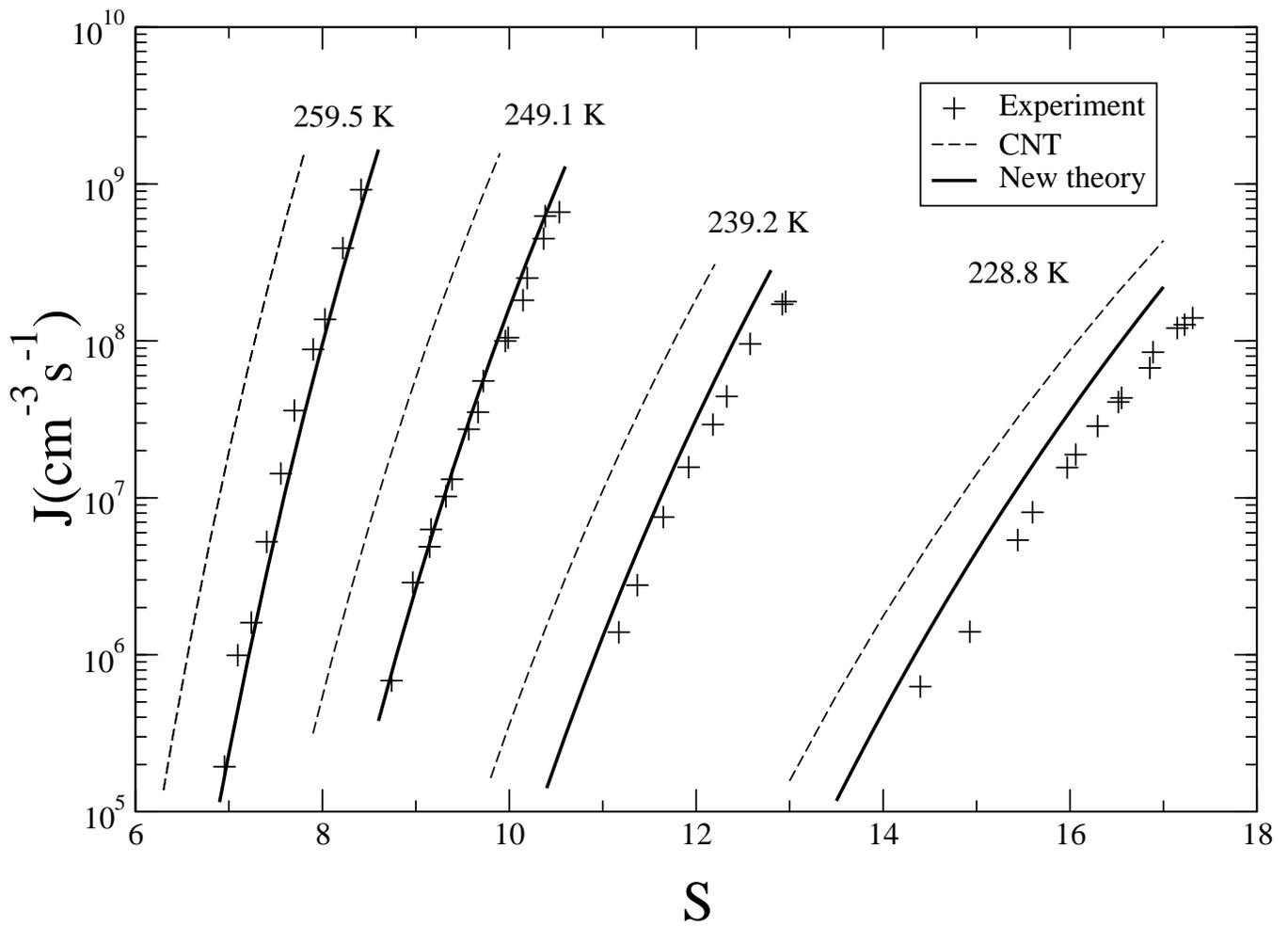}} \par}

\caption{D. Reguera, J.M. Rubí. }
\end{figure}

\end{document}